\documentstyle[aps,epsfig,preprint]{revtex}
\newcommand{\labx}[1] 
{\label{#1}}

\tighten
\begin{document}
\draft
\preprint{SUSX-TH-99-002}
\title{Simulating hot Abelian gauge dynamics}
\author{
A.~Rajantie\thanks{a.k.rajantie@sussex.ac.uk}
and 
M.~Hindmarsh\thanks{m.b.hindmarsh@sussex.ac.uk} 
}
\address{Centre for Theoretical Physics,\\ University of Sussex,\\ 
Brighton BN1 9QJ,\\ UK}
\maketitle
\begin{abstract}
The time evolution of soft modes in a quantum gauge
field theory is to first approximation classical, but the 
equations of motion are non-local. We show how they can be written in
a local and Hamiltonian way in an Abelian theory, and that this
formulation is particularly suitable for numerical simulations.
This makes it possible to simulate numerically non-equilibrium
processes such as the phase transition in the Abelian Higgs model and 
and to study,
for instance, bubble nucleation and defect formation. This would also
help to understand the phase transitions in more complicated gauge theories.
Moreover, we show that the existing analytical results for the
time-evolution in a pure-gauge theory correspond to a special class
of initial conditions and that different initial conditions can lead to
qualitatively different behavior. We compare the results of the
simulations to analytical calculations and find an excellent agreement.
\end{abstract}
\narrowtext
\section{Introduction}
\label{sec:intro}
The recent finding that the electroweak phase transition might be only
a smooth cross\-over~\cite{Kajantie:1997qd} has changed the picture of 
cosmological phase transitions in a fundamental way and shown that theories
with gauge fields can behave very differently from those with only
global symmetries. However, the statement about the smoothness of the
transition refers only to the equilibrium properties, and from the point
of view of cosmology, it is very important to understand also the
time evolution of the transition, in particular, whether the fields
fall out of equilibrium and what its consequences would be. For this,
a calculation scheme is needed for dynamical quantities that
treats the gauge fields correctly and gives the same results in
the special case of thermal equilibrium as the
the dimensional reduction method~\cite{Ginsparg:1980ef,Kajantie:1996dw} 
that was used to show the
smoothness of the electroweak transition. In this paper, we propose
such a scheme.

Instead of the electroweak theory, we will concentrate here on
scalar electrodynamics, i.e.~the Abelian Higgs model.
Not only does the simpler gauge group allow more efficient simulations,
but the theory also contains topological defects, namely 
Nielsen-Olesen vortices~\cite{Nielsen:1973ve}, 
and therefore it can be used to
simulate defect formation in a phase transition~\cite{Kibble:1976sj} and
to test the validity of the present predictions for the produced
defect density~\cite{Zurek:1996sj}
in gauge theories. 

In Ref.~\cite{Vincent:1998cx},
the dynamics of the transition and, in particular, formation of vortices
was simulated classically on a lattice
with field equations derived from the original Lagrangian.
A similar approach had been suggested earlier~\cite{Grigoriev:1988bd}
in the context of electroweak baryogenesis.
This was based on the observation that the quantum distribution
of the long-wavelength modes is essentially classical.
Since the time evolution of a classical
field theory is given by the equations of motion, this approach
allows non-perturbative numerical simulations. For baryogenesis,
the interesting quantity is the hot sphaleron rate (see 
Ref.~\cite{Rubakov:1996vz} and references therein), which is an
equilibrium quantity, but similar methods can also be used to
describe small deviations from equilibrium, such as phase transitions.
After all, the phase of the system is only a property of the long-wavelength
modes, and the
distribution of the high-momentum modes is to a large extent 
independent of it. Therefore also the transition between the phases is
described by the classical long-wavelength modes.

This straightforward approach has the serious problem that
the results depend on the ultraviolet 
cutoff\cite{Bodeker:1995pp,Arnold:1997yb}.
Because of the Rayleigh-Jeans ultraviolet divergences, a classical
continuum field theory cannot actually even be in thermal 
equilibrium -- this was one of the reasons
why quantum mechanics was invented in the first place! 
The divergences depend on the temperature and are absent at zero temperature, 
so they cannot be
cancelled by introducing counterterms as in quantum theories.
Therefore it is not sufficient to
use the classical theory with the Lagrangian of the original
quantum theory, but one has to construct an effective Lagrangian for
the low-momentum (soft) modes by integrating out the high-momentum (hard)
modes. The effective Lagrangian then contains precisely the necessary
(temperature-dependent) counterterms to remove the divergences.

In calculating static properties, i.e.~expectation values of products
of operators measured
at the same time, this procedure is known as dimensional 
reduction\cite{Ginsparg:1980ef}, and leads to a three-dimensional effective
theory, in which the temporal component of the gauge field
has a Debye mass term $m_D\sim gT$. To calculate non-static quantities, the
full four-dimensional effective Lagrangian needs to be constructed
by calculating the so-called hard thermal loops, and 
in gauge theories it
turns out to be non-local~\cite{Silin:1960,Pisarski:1989vd}. 
This makes
computer simulations practically impossible, since one would have to
keep in the memory all the configurations encountered during the time
evolution.

Fortunately, the equations of motion can be written in a local form
by introducing auxiliary fields\cite{Nair:1994xs,Blaizot:1993zk}.
The Abelian version of the resulting system of differential equations is 
known in plasma physics as the Vlasov equation and describes collisionless
electron plasma. Numerical solution of the equations of motion is
still difficult, since the field $W$ depends not only on the coordinate
but also on the momenta of the hard particles it describes, and
therefore the theory becomes essentially five-dimensional.
The traditional method is to approximate it by a 
large number of point particles.
This approach has been used to determine the hot sphaleron rate in the
electroweak theory in Ref.~\cite{Moore:1998sn}. 

In this paper, we show how the theory can be formulated in such a way
that simulations in terms of fields rather than particles becomes
feasible.  One of the problems we have to face is the multiplication of 
the number of degrees of freedom when the hard modes are included.  We show 
that the number of extra degrees of freedom per lattice site must be chosen 
carefully if the simulation of a particular soft mode is to be trusted: 
if one wants to reproduce Landau damping in a mode of momentum 
$k$, naive methods require of order $(kt)^2$  degrees of freedom per lattice site for the 
hard modes.  However, we are able to rewrite the 
effective hard thermal loop improved 
classical equations of motion in such a way that we only need of order 
$kt$ extra degrees of freedom per lattice site.  Our numerical methods 
reproduce known analytic results in pure Abelian gauge theory extremely well, which 
gives us confidence that we will be able to tackle the Abelian Higgs model 
accurately.

The structure of the paper is the following. In Sec.~\ref{sec:kinetic} we
review the hard thermal loop Lagrangian and derive the Hamiltonian 
formulation for the hard modes. In Sec.~\ref{sec:simu} we explain how 
to approximate the system with a finite number of degrees of freedom 
in numerical simulations. In Sec.~\ref{sec:anaresults} we derive some
analytical results, which we compare to our numerical results in 
Sec.~\ref{sec:numerical}. Extending our technique to include the soft
Higgs field is discussed in Sec.~\ref{sec:higgs}, and Sec.~\ref{sec:disc}
contains the conclusions. The paper also has one appendix, in which
the continuum and lattice equations of motion for the Legendre modes
are given explicitly.

\section{Kinetic formulation}
\label{sec:kinetic}
Let us consider scalar electrodynamics at a temperature $T$. The Lagrangian
of the theory is 
\begin{equation}
{\cal L}=-\frac{1}{4}F_{\mu\nu}F^{\mu\nu}+|D_\mu\phi|^2
-m^2|\phi|^2-\lambda|\phi|^4,
\labx{equ:lagr}
\end{equation}
where $D_\mu=\partial_\mu+ieA_\mu$.
We assume that $e\ll 1$ and $\lambda\sim e^2$ and that $m\ll T$ so that
high-temperature approximation can be used.

The phase structure of the theory
was determined in 
Refs.~\cite{Dimopoulos:1997cz,Kajantie:1997hn}. When 
the Higgs self-coupling 
is small, there
is a first-order phase transition, but if 
the self-coupling 
is large enough,
the transition becomes continuous. Unlike in
the electroweak theory, it does not become a smooth 
crossover. In Ref.~\cite{Kajantie:1998zn},
it was pointed out that this can be interpreted as a consequence
of the existence of Nielsen-Olesen vortices. 

To one-loop order, the only non-vanishing contributions to the effective
Lagrangian arise from the two-point diagrams.
The correct procedure
would be to calculate the diagrams in
the original quantum theory and to write down a classical Lagrangian
that gives the same result when the diagrams are calculated on
a lattice~\cite{Bodeker:1995pp}. However, this is 
not possible in practice, since the form of the necessary
effective lattice 
Lagrangian is not known and is presumably very complicated because
the lattice has much less symmetry than the continuum.
In sphaleron rate calculations,
the effective Lagrangian
was taken from the continuum theory and the parameters were fixed by
matching only the static quantities~\cite{Moore:1998sn}. For the simulations
presented in this paper, this problem does not arise, as will be discussed 
later.

In the high-temperature
approximation, the one-loop scalar self-energy
is simply
\begin{equation}
\Sigma=-\left( \frac{e^2}{4}+\frac{\lambda}{3}\right)T^2.
\end{equation}
The photon self-energy~\cite{Kraemmer:1995az} 
is more complicated and if the external momentum is
$K=(\omega,\vec{k})$, it can be written in the form
\begin{equation}
\Pi^{\mu\nu}=\Pi_TP_T^{\mu\nu}+\Pi_LP_L^{\mu\nu},
\end{equation}
where $P_T^{\mu\nu}$ and $P_L^{\mu\nu}$ are the spatially
transverse and longitudinal
projection operators
\begin{equation}
P_T^{ij}=
\frac{k^ik^j}{k^2}-\delta^{ij},\quad P_T^{\mu 0}=0,\quad
P_L^{\mu\nu}=g^{\mu\nu}-\frac{K^\mu K^\nu}{K^2}-P_T^{\mu\nu},
\end{equation}
and
\begin{equation}
\Pi_L=m_D^2\left(1-\frac{\omega^2}{k^2}\right)
\left(1-\frac{\omega}{2k}\ln\frac{\omega+k}{\omega-k}\right),\quad
\Pi_T=\frac{1}{2}\left(m_D^2-\Pi_L\right).
\end{equation}
The Debye mass has the value $m_D^2=\frac{1}{3}e^2T^2+\delta m_D^2$, where
$\delta m_D^2$ is a cutoff-dependent counterterm.

The self-energies can be resummed to a simple effective 
Lagrangian for the soft modes
\begin{eqnarray}
{\cal L}_{\text{HTL}}&=&-\frac{1}{4}F_{\mu\nu}F^{\mu\nu}
-\frac{1}{4}m_D^2
\int\frac{d\Omega}{4\pi}F^{\mu\alpha}
\frac{v_\alpha v^\beta}{(v\cdot\partial)^2} F_{\mu\beta}\nonumber\\
&&+|D_\mu\phi|^2
-m_T^2|\phi|^2-\lambda|\phi|^4,
\labx{equ:lagrHTL}
\end{eqnarray}
where $m_T^2=m^2+(e^2/4+\lambda/3)T^2+\delta m_T^2$, and the integration 
is taken over the unit
sphere of velocities $v=(1,\vec{v})$, $\vec{v}^2=1$.
Note that the form of $m_T^2$ justifies using the high-temperature
approximation even slightly below 
the transition. At tree level, the transition
takes place when $m_T^2=0$, which shows that $m^2\sim -e^2T_c^2$. 
The minimum of the potential is therefore at $v\sim T_c$, and the mass given
by the Higgs mechanism is $m_H^2\sim e^2T_c^2\ll T^2$.

The equations of motion derived from the Lagrangian (\ref{equ:lagrHTL})
are
\begin{mathletters}
\begin{eqnarray}
\partial_\mu F^{\mu\nu}&=&m_D^2\int\frac{d\Omega}{4\pi}
\frac{v^\nu v^i}{v\cdot\partial}E^i-2e{\rm Im}
\phi^*D^\nu\phi,
\labx{equ:nonlA}
\\
D_\mu D^\mu\phi&=&-m_T^2\phi-2\lambda(\phi^*\phi)\phi.
\labx{equ:nonlf}
\end{eqnarray}
\end{mathletters}
The derivative in the denominator in Eq.~(\ref{equ:nonlA}) means that
in order to simulate it numerically, one needs to keep the whole time
evolution of the system in the memory at the same time, which is impossible.
The form of the
equation of motion
(\ref{equ:nonlf}) of the scalar field is much simpler, and actually the only
effect of the hard modes is the modified mass term.
The scalar field can therefore be simulated with standard methods and
we will neglect it from now on, concentrating only on the gauge field.
We will refer to this theory as the pure gauge theory although
it contains the contribution from the hard scalar modes.

The non-locality problem of Eq.~(\ref{equ:nonlA})
can be solved by introducing new 
fields\cite{Nair:1994xs,Blaizot:1993zk}. We follow Ref.\cite{Blaizot:1993zk} 
and add the field $W(x,\vec{v})$,
which satisfies the equation of motion
\begin{equation}
(v\cdot\partial)W(x,\vec{v})=\vec{v}\cdot\vec{E}(x),
\labx{equ:eomW}
\end{equation}
and replacing Eq.~(\ref{equ:nonlA}) with
\begin{equation}
\partial_\mu F^{\mu\nu}=j_W^\nu(x)+j^\nu(x),
\labx{equ:eomA}
\end{equation}
where
\begin{equation}
j_W^\nu(x)=m_D^2\int\frac{d\Omega}{4\pi}v^\nu W(x,\vec{v}),
\labx{equ:currW}
\end{equation}
and $j^\nu(x)$ denotes the current due to the scalar field and any
external currents.

The system of equations (\ref{equ:eomW}), (\ref{equ:eomA}) is 
a special case of
what is known as the
Vlasov equation in plasma physics, where it describes collisionless
electron plasma. The field $W(x,\vec{v})$ gives 
the deviation of the density of
hard particles of velocity $\vec{v}$ from their equilibrium distribution.

In addition to the equations of motion, we also have to specify the
initial conditions, and since we would like to have the system initially
in a thermal equilibrium, they should be given by the equilibrium
distribution, i.e.~we should have a large number of initial configurations
with a Boltzmann distribution 
$\propto\exp(-\beta H)$, where $H$ is the Hamiltonian. However, the equation
of motion (\ref{equ:eomW}) for the $W$ field is not of canonical form,
which makes it difficult to find the correct Hamiltonian.
The suggestion of\cite{Iancu:1997md} was to use the
conserved quantity
\begin{equation}
H=\frac{1}{2}\int d^3x\left\{\vec{E}^2+\vec{B}^2+
m_D^2\int\frac{d\Omega}{4\pi}W(x,\vec{v})^2
\right\}.
\labx{equ:hamW}
\end{equation}

In the simulations, one has to approximate the system with a finite
number of degrees of freedom. Normally, this is done by discretizing the
space to a finite lattice, but in our case, the field $W$ also depends on a
two-dimensional continuous parameter $\vec{v}$. There are various ways to
handle it in simulations. In Refs.~\cite{Moore:1998sn,Hu:1997sf}, $W$ was 
approximated with a large number of particles. The other straightforward
options are approximating the velocity integral with a discrete set
of velocities on the unit sphere, or expanding $W$ in terms of
spherical harmonics. However, as we will show next, at least in the Abelian
case it pays to simplify the problem a bit first.

Let us define in Fourier space
\widetext
\begin{eqnarray}
\vec{f}(\omega,\vec{k},z)&=&
im_D\sqrt{\frac{2z^2}{1-z^2}}\int\frac{d\Omega}{4\pi}
\frac{\vec{v}\times\vec{k}}{(\vec{v}\cdot\vec{k})^2}
W(\omega,\vec{k},\vec{v})\left(
\delta(\vec{v}\cdot\hat{k}-z)+\delta(\vec{v}\cdot\hat{k}+z)
\right)\nonumber\\&&
-\frac{im_D}{k^2}\sqrt{\frac{1-z^2}{2z^2}}\vec{k}\times\vec{A},\nonumber\\
\theta(\omega,\vec{k},z)&=&im_D\int\frac{d\Omega}{4\pi}
\frac{1}{\vec{v}\cdot\vec{k}}
W(\omega,\vec{k},\vec{v})\left(
\delta(\vec{v}\cdot\hat{k}-z)+\delta(\vec{v}\cdot\hat{k}+z)
\right)
+\frac{im_D}{k^2}\vec{k}\cdot\vec{A}.
\labx{equ:defft}
\end{eqnarray}
Then the equations of motion in the temporal gauge $A_0=0$ are
\begin{mathletters}
\label{equ:formulation}
\begin{eqnarray}
\partial_0^2{\vec{f}}(z)&=&z^2\vec{\nabla}^2\vec{f}+m_Dz\sqrt\frac{1-z^2}{2}
\vec\nabla\times\vec{A},\\
\partial_0^2{\theta}(z)&=&z^2\vec{\nabla}\cdot\left(
\vec{\nabla}\theta-m_D\vec{A}\right),\\
\partial_0^2{\vec{A}}&=&-\vec\nabla\times\vec\nabla\times\vec{A}
+m_D\int_0^1dzz^2\left(
\vec\nabla\theta-m_D\vec{A}+
\sqrt{\frac{1-z^2}{2z^2}}\vec\nabla\times\vec{f}
\right)+\vec{j}.
\end{eqnarray}
\end{mathletters}
One advantage of this formulation is that these equations of motion are
canonical and the corresponding Hamiltonian is
\begin{eqnarray}
H&=&\frac{1}{2}\int d^3x\int_0^1dz\Biggl[
\vec{E}^2+(\vec\nabla\times\vec{A})^2+\vec{F}^2+{\Pi}^2+
z^2(\vec\nabla\times\vec{f})^2+z^2(\vec{\nabla}\theta-m_D\vec{A})^2
\nonumber\\&&
-m_Dz\sqrt\frac{1-z^2}{2}\vec{f}\cdot\vec\nabla\times\vec{A}
\Biggr],
\labx{equ:hamf}
\end{eqnarray}
\narrowtext
where $\vec{F}=\partial_0{\vec{f}}$ and $\Pi=\partial_0\theta$ 
are the canonical momenta of $\vec{f}$ and $\theta$,
respectively. We also need two extra conditions, namely the transverseness of
$\vec{f}$ and Gauss's law
\begin{eqnarray}
\vec\nabla\cdot\vec{f}&=&\vec\nabla\cdot\vec{F}=0,\nonumber\\
\vec\nabla\cdot\vec{E}&=&-m_D\int_0^1dz\Pi(z).
\labx{equ:Gauss}
\end{eqnarray}

By reformulating the degrees of freedom, we have gained two important
advantages. First, $\vec{f}$ depends only on one internal coordinate whereas
$W$ depends on two, and second, since we know the Hamiltonian 
(\ref{equ:hamf}), we know the correct distribution of the initial
configurations.

\section{Simulations}
\label{sec:simu}
We still have to approximate the $z$-dependence of the hard modes with a
finite number of degrees of freedom, and we will use Legendre polynomials
for that.
We define
\begin{eqnarray}
\vec{f}^{(n)}&=&\int_0^1dzz\sqrt\frac{1-z^2}{2}P_{2n}(z)\vec{f}(z),
\nonumber\\
\theta^{(n)}&=&\int_0^1dzz^2P_{2n}(z)\theta(z),
\labx{equ:defLeg}
\end{eqnarray}
where $P_{n}(z)$ is the $n$th Legendre polynomial. The equations
of motion for different Legendre modes are given in Eq.~(\ref{equ:eomLeg})
in the appendix. It is also shown in the appendix that the approximation
can be trusted if the simulation time is less than
\begin{equation}
t_0\approx 2N_{\text{max}}/k. 
\labx{equ:reltime}
\end{equation}

On the lattice, the field $\theta$ is defined on lattice sites, while
$\vec{f}$ and the gauge field $\vec{A}$ are defined on links between 
the sites in
such a way that the invariance under time-independent gauge
transformations is preserved. Note that the field $\theta^{(0)}$ is
not gauge invariant by itself, but the combination 
$\vec\nabla\theta^{(0)}-\frac{1}{3}m_D\vec{A}$ is.
The canonical momenta are defined at half-way between the time steps so
that the value of, say, $\vec{f}$ at time $t+\delta t$ is determined
from the fields at time $t$ and the momenta at time $t+\delta t/2$.
In this way, the time reversal invariance of the continuum theory is 
preserved. The lattice equations of motion are given in 
Eq.~(\ref{equ:eomLatt}) in the appendix.

The aim of the simulations described here was mainly to
test the simulation code by comparing it with known analytical results.
Therefore we neither averaged over thermal initial conditions nor
included the scalar field.
The equations of motion for the pure gauge theory
are linear and can therefore in principle be solved analytically.
It also means that the modes with different momenta do not interact
with each other, and to simulate a mode with momentum along, say, the
$z$ axis, we can use a lattice with $1\times 1\times N_z$ sites. This
makes the simulations very simple. Moreover, the pure gauge theory is
ultraviolet finite, so no mass counterterms are needed.
In particular, $\delta m_D^2$ and $\delta m_T^2$ introduced in
Sec.~\ref{sec:kinetic} vanish.

\section{Analytical Results}
\label{sec:anaresults}
Before specializing to particular solutions, we first discuss the 
finite temperature propagator for the pure Abelian theory.
Due to linearity, it is sufficient to consider a single
Fourier mode at a time. Let the momentum be $\vec{k}$, and, for simplicity,
let us consider just the transverse part of the propagator $G(t,\vec{x})$.
Its Fourier transform 
has a non-analytic form
\begin{equation}
\frac{1}{G(\omega,\vec{k})}=
-\omega^2+k^2+\frac{m_D^2}{4}\left[
2\frac{\omega^2}{k^2}+\frac{\omega}{k}\frac{k^2-\omega^2}{k^2}
\ln\frac{\omega+k}{\omega-k}
\right].
\labx{equ:prop}
\end{equation}
The analytic structure of the propagator is shown in Fig.~\ref{fig:prop}.
The propagator has two oscillatory poles on the real axis at 
$\omega^2_p\approx k^2+\frac{1}{3}m_D^2$, but also a branch cut from
$\omega=-k$ to $\omega=k$.
The original equation of motion (\ref{equ:nonlA}) implies that the branch
cut must be taken along the real axis. There are no other singularities
on this physical Riemann sheet, but if one continues the propagator
analytically from the upper half-plane through the branch cut,
one finds a pole at
\begin{equation}
\omega=-i\gamma_L \approx -i\frac{4k^3}{\pi m_D^2}.
\end{equation}
Likewise, continuing analytically from the lower half-plane through the
cut reveals a pole at $\omega=i\gamma_L$.

The solution to the inhomogeneous equation with an external 
source 
$\vec{j}(\omega)$ is, assuming that the current is transverse 
and that the fields vanish at $t=-\infty$,
\begin{equation}
\vec{A}(t,\vec{k})=
\int_{-\infty+i\epsilon}^{\infty+i\epsilon}\frac{d\omega}{2\pi}
e^{-i\omega t} G(\omega)\vec{j}(\omega).
\labx{equ:formalsol}
\end{equation}
If $t>0$, the integral can be transformed into a contour integral by
closing it on the lower half-plane.
This integral has three pieces: two contributions from 
the poles and one from the integral around the branch cut.
The result of the integration around 
the cut is sensitive to the functional form of $\vec{j}(\omega)$, a 
reflection of the non-locality of Eq.~(\ref{equ:nonlA}): the behavior of 
the system depends on its whole history.

For example, Boyanovsky et al.~\cite{Boyanovsky:1998pg} discussed a situation
in which an inhomogeneous initial configuration is set up for the
soft fields $\vec{A}$ and $\vec{E}$, and calculated its relaxation 
to the equilibrium at asymptotically
long times. Still keeping the assumption of transverseness, 
the solution was given as a function
of $\vec{A}$ and $\vec{E}$ at time $t=0$,
\begin{eqnarray}
\vec{A}(t,\vec{k})&\approx&
\vec{A}(0,\vec{k})
\left[Z[T]\cos\omega_pt-\frac{4}{m_D^2}\frac{\cos kt}{t^2}\right]
-\vec{E}(0,\vec{k})
\left[Z[T]\frac{\sin\omega_pt}{\omega_p}-
\frac{4}{m_D^2}\frac{\sin kt}{kt^2}\right],
\labx{equ:boyaresult}
\end{eqnarray}
where 
\begin{equation}
Z[T]=-\left(\frac{\partial G^{-1}}
{\partial \omega^2}\right)^{-1}_{\omega=\omega_p}.
\end{equation}
Eq. 
(\ref{equ:boyaresult}) can be seen to correspond to
$\vec{j}(\omega)=-i\omega\vec{A}(0)-\vec{E}(0).$
As they emphasize, the dominant contribution does not come from
the smallest frequencies but those near the end points of the branch cut.

With a different 
choice of $\vec{j}$, qualitatively different solutions
can be found. Suppose that $\vec{j}(t)$ is increased very
slowly from zero to a finite value and then suddenly switched off,
i.e.~$\vec{j}(t)=\vec{j}_0e^{\gamma_0 t}\Theta(-t)$, 
where $\gamma_0$ is eventually
taken to zero and $\Theta(t)$ is the step function.
The Fourier transform $\vec{j}(\omega)=\vec{j}_0/(\gamma_0+i\omega)$
is peaked around the origin, and therefore the integral around the branch 
cut is dominated by the integrand near $\omega=0$. Hence, expanding 
the propagator around $\omega=i\epsilon$ and $\omega=-i\epsilon$ on
the upper and lower half-planes, respectively,  we may write 
\begin{equation}
\left.\vec{A}(t,\vec{k})\right|_{\rm cut}\approx
\vec{j}_0\int_{-k}^k\frac{d\omega}{2\pi}\left ( e^{-i\omega t}
\frac{1}{\gamma_0+i\omega}\frac{1}{\gamma_L - i\omega} + 
e^{i\omega t}
\frac{1}{\gamma_0-i\omega}\frac{1}{\gamma_L - i\omega}
\right)
\frac{\gamma_L}{k^2}.
\end{equation}
The 
limits of the integral may be taken to infinity, and the contour closed 
in either the upper or the lower half-plane depending on the sign of the 
exponent. The second term has no poles in the upper half-plane and hence 
vanishes, while the first term gets a contribution from the pole at 
$\omega = -i\gamma_L$. 
Thus we find an exponentially damping solution
\begin{equation}
\left.\vec{A}(t,\vec{k})\right|_{\rm cut}
\approx -\frac{\vec{j}_0}{k^2} e^{-\gamma_Lt}.
\end{equation}
This exponential decay is known as Landau damping, and it is important to
notice that it did not really come from integrating around a 
pole of the propagator 
(\ref{equ:prop}). If it had, then the equations of motion would
either have to break time reversal invariance, which they do not do,
or there would have to be a corresponding exponentially growing
solution, thus making the system unstable.

For completeness, we also calculate the contribution coming from the
poles at $\omega=\pm\omega_p$. The full result is then
\begin{equation}
\vec{A}(t,\vec{k})\approx 
-\vec{j}_0\left(
Z[T]
\frac{\cos\omega_pt}
{\omega_p^2}
+
\frac{1}{k^2} e^{-\gamma_Lt}
\right).
\labx{equ:landauasymp}
\end{equation}

\section{Numerical results}
\label{sec:numerical}
\begin{figure}
\begin{center}
\epsfig{file=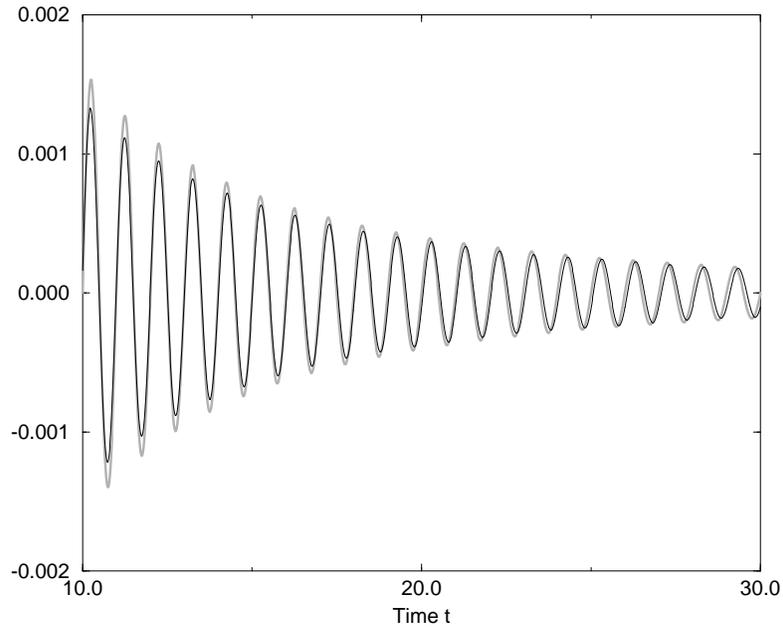,width=12cm}
\end{center}
\caption{The comparison of Eq.~(\ref{equ:boyaresult}) (gray line)
to the numerical 
result (black line) with $k=m_D=2\pi$. 
The constant-amplitude oscillation has been subtracted.}
\label{fig:boyacmp}
\end{figure}

In order to numerically 
reproduce the results (\ref{equ:boyaresult}), (\ref{equ:landauasymp}),
we have to specify
the correct initial conditions in terms of $\vec{f}$ and $\theta$.
The result (\ref{equ:boyaresult}) is given simply by $\vec{f}(0)=\theta(0)=0$.
For the soft modes, we chose 
$\vec{A}(0)=0$, $\vec{E}(0)=10\hat{y}\sin \vec{k}\cdot\vec{x}$. 
To see the power-law damping most 
clearly, $m_D$ must be relatively small, and we chose $m_D=2\pi$, 
$\vec{k}=2\pi\hat{x}$ in
our units. The lattice size was $20\times 1\times 1$, lattice spacing 
$a=0.05$ and the time step $\delta t=0.01$. We used $N_{\rm max}=200$
so that according to Eq.~(\ref{equ:reltime})
the results should be reliable when $t\lesssim 60$. 
Eq.~(\ref{equ:boyaresult}) becomes
\begin{equation}
\vec{A}(t,\vec{k})\approx
\left(-1.236\sin 7.451x+0.1613\frac{\sin 2\pi t}{t^2}\right)\hat{y}.
\labx{equ:boyanum}
\end{equation}
We added to the numerical result a constant-amplitude part
$\alpha\sin\beta t$ and determined the parameters $\alpha$ and
$\beta$ from the condition that the remaining part agrees with the
decaying part of Eq.~(\ref{equ:boyanum}). The result with $\alpha=1.24008$
and $\beta=7.43024$ is shown in Fig.~\ref{fig:boyacmp}. The discrepancy 
between the numerical and analytical results is less than one percent.

\begin{figure}
\begin{center}
\epsfig{file=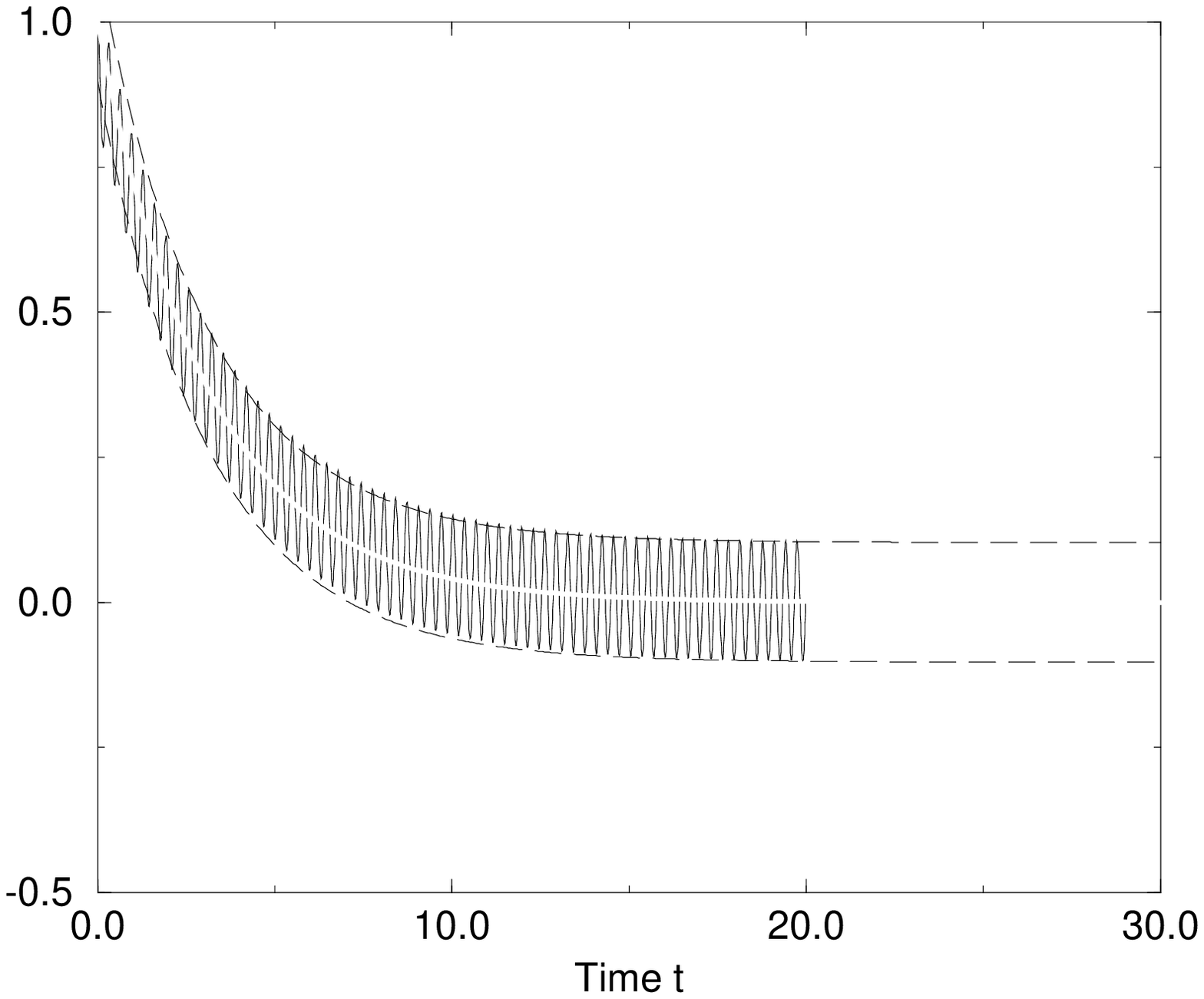,width=8cm}\epsfig{file=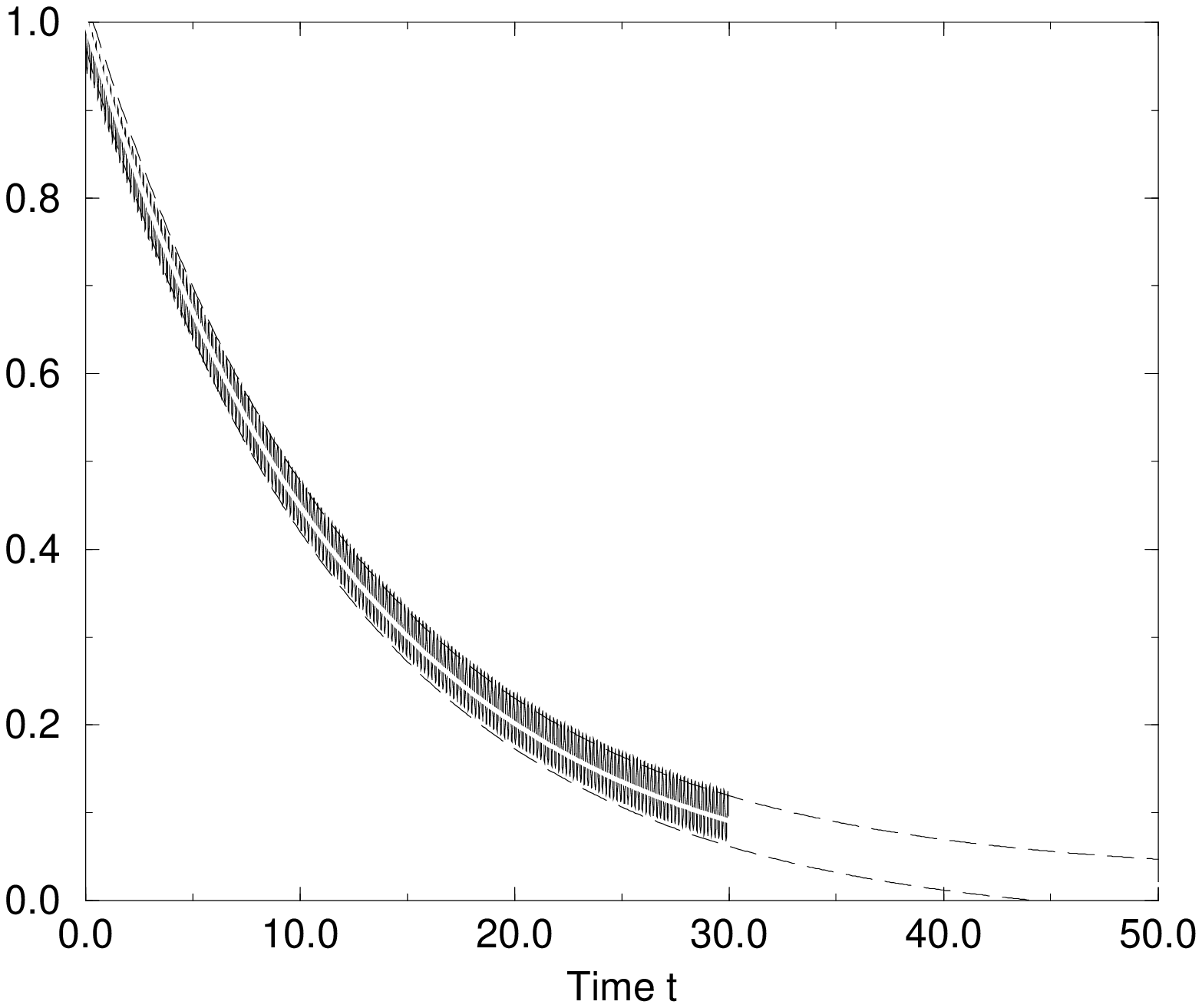,width=8cm}
\end{center}
\caption{Results of the Landau damping simulations with $k=2\pi$ and
$m_D=10\pi$ (left)
and $m_D=20\pi$ (right). The white line is the predicted Landau damping rate
$\gamma_L$ 
and the dashed lines show the envelope of the analytical result 
(\ref{equ:landauasymp}).}
\label{fig:landaucmp}
\end{figure}

The initial conditions that correspond to Eq.~(\ref{equ:landauasymp})
are those in which the first and second time derivatives of the fields
vanish. If we choose $\vec{j}_0$ such that 
$\vec{A}(0,\vec{x})=\hat{y}\sin \vec{k}\cdot\vec{x}$, 
then Eqs.~(\ref{equ:defft}) and 
(\ref{equ:defLeg})
imply that 
\begin{eqnarray}
\vec{f}^{(0)}(0,\vec{x})&=&\frac{m_D}{3k}\hat{z}
\cos \vec{k}\cdot\vec{x},\nonumber\\
\vec{f}^{(1)}(0,\vec{x})&=&-\frac{m_D}{15k}\hat{z}
\cos\vec{k}\cdot\vec{x},\nonumber\\
\vec{f}^{(n>1)}(0,\vec{x})&=&0,\quad\theta^{(n)}(0,\vec{x})=0.
\end{eqnarray}
The exponential damping is seen most clearly if $m_D\gg k$, and we chose
two different values $m_D=10\pi$ and $m_D=20\pi$ while we took again
$\vec{k}=2\pi\hat{x}$.
Again, the lattice size was $20\times 1\times 1$, the lattice spacing
$a=0.05$, the time step $\delta t=0.01$, and $N_{\rm max}=200$. The results
of the simulations are shown in Fig.~\ref{fig:landaucmp}. The predicted decay 
rates are $\gamma_L=0.32$ and $\gamma_L=0.08$ and the amplitudes of the
oscillation $A_{\rm osc}\approx 0.103$ and $A_{\rm osc}\approx0.029$.
As can be seen from Fig.~\ref{fig:landaucmp}, these agree very well with the
numerical results.

\begin{figure}
\begin{center}
\epsfig{file=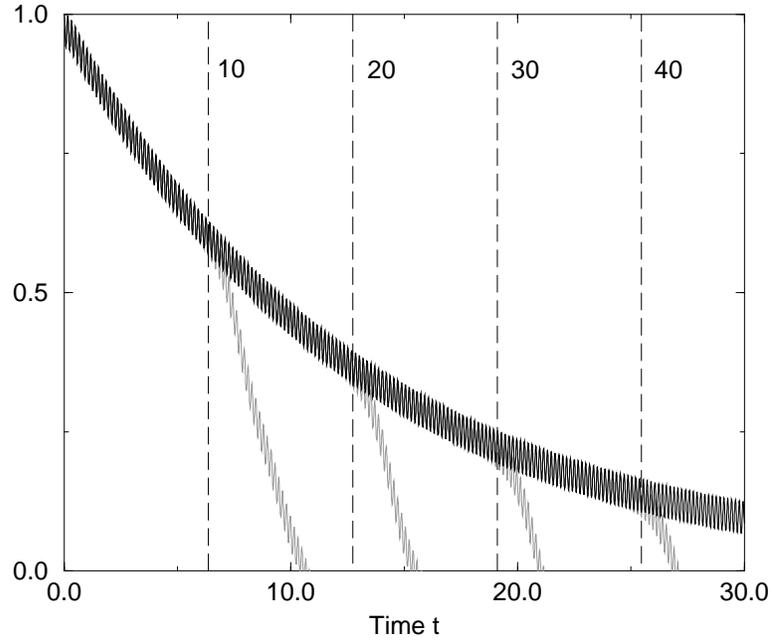,width=12cm}
\end{center}
\caption{Landau damping simulations with $N_{\rm max}=10\ldots 40$
($k=2\pi$, $m_D=20\pi$). The
vertical dashed lines show the analytical prediction 
$t\approx 4N_{\rm max}/k$ for the time
when the approximation
should break down. The black curve is the result with $N_{\rm max}=200$
and the gray curves correspond to $N_{\rm max}=10,20,30,40$ from left to 
right.}
\label{fig:testNmax}
\end{figure}

We also carried out simulations with with different values of $N_{\rm max}$
to find out at what time the approximation breaks down and to 
test the estimate (\ref{equ:reltime}). We used $m_D=20\pi$ and the values 
for the other parameters were as before. 
Since all the Legendre modes with $n>1$ are
initially zero, we expect the first errors to occur at time 
$t\approx 2t_0\approx 4N_{\rm max}/k$. 
The result is shown in Fig.~\ref{fig:testNmax}
for various values of $N_{\rm max}$ and confirms our estimate.
The number of extra scalar degrees of freedom needed for simulating
with any particular value of
$N_{\rm max}$ is $8N_{\rm max}$.

\begin{figure}
\begin{center}
\epsfig{file=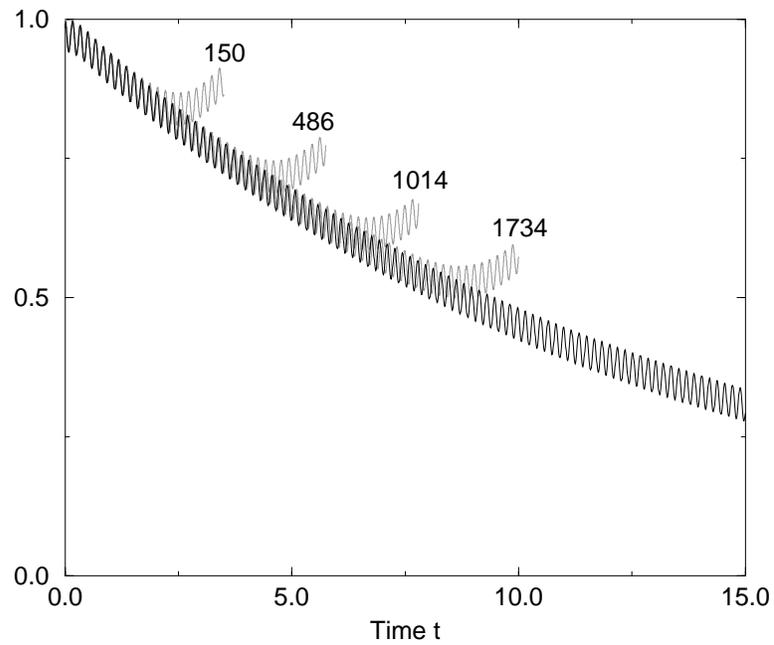,width=12cm}
\end{center}
\caption{Landau damping simulations with $k=2\pi$, $m_D=20\pi$
in a formulation in which a large
number of points on the unit sphere of velocities is used to approximate
the integral. The values of $N_v$ were (from left to right)
2, 4, 6 and 8 and the corresponding numbers of degrees of freedom
are shown in the plot.}
\label{fig:testvkoko}
\end{figure}

In order to compare the efficiency of the Legendre polynomial formulation
with other approaches, we carried out
the same simulations with a more straightforward method. We chose a
large number of points on the unit sphere to represent different velocities
and used them to simulate the pair of equations (\ref{equ:eomW}), 
(\ref{equ:eomA}). More precisely, the different velocities were
\begin{equation}
\vec{v}=\frac{(N_v+\frac{1}{2},
n_y,n_z)}{\sqrt{(N_v+\frac{1}{2})^2+n_y^2+n_z^2}},
\quad -N_v\le n_i\le N_v,
\end{equation}
and those obtained from that with reflections or rotations of $\pi/2$.
The parameters were the same as in Fig.~\ref{fig:testNmax}.
The values of $N_v$ used ranged from 2 to 8, and the corresponding
number of extra degrees of freedom is $6(2N_v+1)^2$.
The results in Fig.~\ref{fig:testvkoko} show clearly
that the number of different velocities becomes quickly prohibitive
when the simulation time is increased. 
If we assume that, as with the Legendre polynomials, the particularly
smooth initial conditions result in a factor of two in the reliable
simulation time, we can 
estimate that generally
a simulation will be reliable if $t\lesssim \pi N_v/k$.  

This estimate can also be reached analytically. Since the time evolution is
a superposition of oscillations, the reliable simulation time is
$t_0\approx \pi/\omega_0$, where $\omega_0$ is the smallest frequency,
i.e.~the pole of the propagator that is nearest to the origin.
The transverse self-energy $\Pi_T(\omega)$ diverges if 
$\omega=\vec{k}\cdot\vec{v}$ for any velocity
$\vec{v}$. 
The smallest pole of the propagator is at a frequency $\omega_0$, which is
smaller than $\min\{\vec{k}\cdot\vec{v}\}$, but of the same
order of magnitude. Generally $\min\{\vec{k}\cdot\vec{v}\}
\approx k/N_v$, leading to the previous estimate.


Since expansion of $W$ in 
Eq.~(\ref{equ:eomW}) in terms of spherical harmonics $Y^m_n$
is essentially equivalent to the expansion of $\vec{f}$ and $\theta$
in terms of Legendre polynomials, we can also estimate the effectiveness
of that approach. In order to reach the same simulation time,
the highest $n$ used should be $2N_{\rm max}$.
The number of extra degrees of freedom would then be
$(2N_{\rm max}+1)^2$, which again increases much faster than in
the Legendre polynomial approach.
Our estimates of the reliable simulation times
in the different approaches have been summarized in table~\ref{tab:simtimes}.
They show that the Legendre polynomial formulation is the most
economical one.
\footnote{
The analogous number of degrees of freedom in the particle
method used by Moore et al.~\cite{Moore:1998sn} is $6\langle n\rangle$.
The value of $\langle n\rangle$ they were using varied between 17 and 120,
but they did not carry out this kind of systematic analysis
of the corresponding reliable simulation time. However,
the intrinsic randomness of the method seems to imply that it does not
reproduce the correct
behavior as accurately at short times as the methods discussed here.}

\section{Higgs model}
\label{sec:higgs}
The pure gauge theory that has been discussed so far is an ideal
way to test the formalism, since it is linear and 
one can in principle solve it exactly. In the previous section, we have
shown that the numerical results agree with the analytical calculations
and we were even able to estimate the number $N_{\rm max}$ of Legendre
modes needed to ensure that the simulation is reliable. 

It is very easy to transform the system to a very non-trivial one simply
by adding a Higgs field. While this makes it impossible to solve
the equations of motion analytically, from the point of view of numerical
simulations it means only adding one more field. Still, the
resulting theory has a complicated structure with a phase transition and
topological defects. 

As was shown in Sec.~\ref{sec:kinetic}, the weak-coupling condition
$e\ll 1$ and $\lambda\sim e^2$ implies that the hard thermal loop
approximation is also valid near the transition in the broken phase.
The reason is that the distribution of the hard modes is independent
of the phase.
Furthermore, when the phase transition takes place in
a finite time, only the soft modes fall out of equilibrium, and therefore
the classical description remains valid, even in such non-equilibrium
processes.

The above suggests that the classical formalism discussed in this paper
can be used to simulate non-equilibrium dynamics of the phase
transition in hot scalar electrodynamics.
In the cosmological setting the transition takes place when the universe
cools as it
expands, but in a simulation some other way to change the temperature
is needed. A straightforward way of doing this is preparing an initial
ensemble with a higher temperature for the soft modes than for the hard modes.
A large number of initial configurations would then be taken from this 
ensemble and evolved in time. When the soft and hard modes interact,
the temperature of the soft modes decreases and they 
undergo a phase transition. 

While this kind of an instantaneous quench in the temperature is not
very realistic, it would still be a first approximation to the
true cosmological phase transition. These simulations would
answer many interesting questions about the dynamics of the phase
transition, for instance the density of topological defects created.

Certainly, the hard thermal loop
approximation we have used
is only valid at relatively
short times.
We have neglected the collisions of the hard
particles and at long times 
$t\gtrsim 1/e^4T$
they would have an essential contribution
to the time evolution~\cite{Blaizot:1999xk}. 
If the coupling is weak, as we assume, this is not a serious problem.
The other problem is to remove the effect of the ultraviolet lattice
modes as discussed in Sec.~\ref{sec:kinetic}.
The fact that the simulations 
of \cite{Vincent:1998cx} showed no signs of a Rayleigh-Jeans catastrophe 
developing over the course of the simulation
seems to indicate that the effect can safely be neglected.

Phase transitions and other non-equilibrium processes can be simulated
also in non-Abelian theories using the Vlasov equation (\ref{equ:eomW})
to describe the hard modes. 
It would be important to know, whether a formulation analogous to
Eq.~(\ref{equ:formulation}) with only one internal
coordinate is also possible in non-Abelian theories, since
that would reduce the need of computational power drastically.
Since the derivative in Eq.~(\ref{equ:eomW})
is replaced with a covariant derivative,
operating in the momentum space becomes more complicated and
the same derivation cannot be used.

\section{Conclusions}
\label{sec:disc}
%
In this paper, we have studied hot Abelian gauge field theory in
the hard thermal loop approximation. Starting from the local kinetic
formulation (\ref{equ:eomW}), (\ref{equ:eomA}), we were able to 
reformulate the degrees of freedom in a more economical way. 
The equations of motion in the new formulation are
canonical and we found the explicit form of the Hamiltonian 
(\ref{equ:hamf}). 

The pure gauge theory discussed in this paper is linear and can therefore
be solved analytically, at least in principle. We pointed out that
because of non-locality it is not sufficient to specify the initial 
conditions for the soft modes only. In a sense, one will have to specify
the whole history of the system in order to calculate its behavior in
the future. Taking this effect into account modifies the existing
results~\cite{Boyanovsky:1998pg} and leads to exponential damping
(\ref{equ:landauasymp})
with the Landau damping rate.

In order to simulate the system, we approximated the functions
$\vec{f}$, $\theta$ with a finite number of Legendre polynomials. The 
simulations reproduced the analytical results, and the number of
degrees of freedom needed to describe the hard modes
was much smaller than in other possible approaches. 
This shows that the
formulation presented in this paper is very well suited for numerical
simulations.

While the pure gauge theory is trivial in the sense that it can be solved
analytically, including the soft Higgs field makes analytical calculations
essentially impossible. On the other hand, it is very simple to add it to
numerical simulations. Leaving possible ultraviolet problems aside,
that would give a way to study non-perturbatively the 
non-equilibrium dynamics of
a phase transition in a gauge field theory.

\acknowledgments
AR would like to thank D.~B\"odeker, D.~Boyanovsky, E.~Iancu, M.~Laine, 
G.D.~Moore and
K.~Rummukainen for useful discussions. This work was supported by 
PPARC grant GR/L56305.
AR was partly supported by the
University of Helsinki.

\appendix
\section{Equations of motion for Legendre modes}
The equations of motion for the Legendre modes defined in 
Eq.~(\ref{equ:defLeg}) are
\widetext
\begin{eqnarray}
\partial_0^2{\vec{f}}^{(n)}&=&
C^+_n\vec\nabla^2\vec{f}^{(n+1)}
+C^0_n
\vec\nabla^2\vec{f}^{(n)}
+C^-_n\vec\nabla^2\vec{f}^{(n-1)}\nonumber\\
&&+m_D\vec\nabla\times\vec{A}
\left(\frac{1}{15}\delta_{n,0}+\frac{1}{105}\delta_{n,1}
-\frac{4}{315}\delta_{n,2}
\right),\nonumber\\
\partial_0^2{\theta}^{(n)}&=&
C^+_n\vec\nabla^2\theta^{(n+1)}
+C^0_n
\vec\nabla^2\theta^{(n)}
+C^-_n\vec\nabla^2\theta^{(n-1)}\nonumber\\
&&-m_D\vec\nabla\cdot\vec{A}
\left(\frac{1}{5}\delta_{n,0}+\frac{4}{35}\delta_{n,1}
+\frac{8}{315}\delta_{n,2}
\right),\nonumber\\
\partial_0^2{\vec{A}}&=&-\vec\nabla\times\vec\nabla\times\vec{A}
-\frac{1}{3}m_D^2\vec{A}
+m_D\left(
\vec\nabla\theta^{(0)}+
\vec\nabla\times\vec{f}^{(0)}
\right),
\labx{equ:eomLeg}
\end{eqnarray}
where
\begin{equation}
C^+_n=\frac{(2n+1)(2n+2)}{(4n+1)(4n+3)},\quad
C^0_n=\frac{1}{4n+1}\left(\frac{(2n+1)^2}{4n+3}+\frac{4n^2}{4n-1}\right),\quad
C^-_n=\frac{2n(2n-1)}{(4n+1)(4n-1)}.
\end{equation}

\narrowtext
Approximating an infinitely many degrees of freedom with a finite number
gives necessarily rise to errors. Suppose we only take $N_{\text{max}}$ lowest
Legendre modes into account. If $n\gg 1$, then the equation of motion
for the Fourier mode with momentum $k$ of $\theta^{(n)}$ is simply
\begin{equation}
\partial_0^2{\theta}^{(n)}=
-\frac{1}{4}k^2\left(\theta^{(n+1)}+2\theta^{(n)}+\theta^{(n-1)}
\right).
\end{equation}
By writing $\theta^{(n)}=(-1)^{n}\tilde\theta^{(n)}$, we notice
that this is precisely a discretized version of a wave equation, with
waves propagating at speed $c=k/2$. The same applies to $\vec{f}$ as well.
Since in the approximation, errors
only occur at $n=N_{\text{max}}$, we can estimate that the approximation
works as long as the error has not had time to propagate to $n=0$, which
is the only mode that is coupled to the observable soft modes.
Thus, for a mode with momentum $k$, the approximation is reliable
as long as the simulated time is smaller than 
\begin{equation}
t_0\approx 2N_{\text{max}}/k. 
\eqnum{\ref{equ:reltime}}
\end{equation}
Therefore, if we want to measure
correlators with time separation $t_{\text{max}}$, we need to have
$N_{\text{max}}\gtrsim t_{\text{max}}/2a$, where $a$ is the lattice spacing.

For completeness, we give here the lattice equations of motion
\begin{eqnarray}
E_i(t+\frac{\delta t}{2},\vec{x})&=&E_i(t-\frac{\delta t}{2},\vec{x})\nonumber\\&&
+\delta t\left\{
\frac{1}{a^2}\Delta^-_jA_{ij}(t,\vec{x})
+\frac{1}{3}m^2_DA_i(t,\vec{x})
-\frac{m_D}{a}\left[
\Delta^+_i\theta^{(0)}
+\epsilon_{ijk}\Delta^-_j
f_k^{(0)}(t,\vec{x})
\right]
\right\},\nonumber\\
F_i^{(n)}(t+\frac{\delta t}{2},\vec{x})&=&F_i^{(n)}(t-\frac{\delta t}{2}
,\vec{x})\nonumber\\&&
+\delta t\Biggl\{
C_n^+\tilde{\nabla}^2f_i^{(n+1)}(t,\vec{x})+
C_n^0\tilde{\nabla}^2f_i^{(n)}(t,\vec{x})+
C_n^-\tilde{\nabla}^2f_i^{(n-1)}(t,\vec{x})\nonumber\\&&
+\frac{m_D}{a}\left(\frac{1}{15}\delta_{n,0}+\frac{1}{105}\delta_{n,1}
-\frac{4}{315}\delta_{n,2}\right)\epsilon_{ijk}\Delta^+_jA_k(t,\vec{x})
\Biggr\},\nonumber\\
\Pi^{(n)}(t+\frac{\delta t}{2},\vec{x})&=&\Pi^{(n)}(t-\frac{\delta t}{2}
,\vec{x})\nonumber\\&&
+\delta t\Biggl\{
C_n^+\tilde{\nabla}^2\theta^{(n+1)}(t,\vec{x})+
C_n^0\tilde{\nabla}^2\theta^{(n)}(t,\vec{x})+
C_n^-\tilde{\nabla}^2\theta^{(n-1)}(t,\vec{x})\nonumber\\&&
-\frac{m_D}{a}
\left(\frac{1}{5}\delta_{n,0}+\frac{4}{35}\delta_{n,1}
+\frac{8}{315}\delta_{n,2}
\right)\Delta_i^-A_i(t,\vec{x})\Biggr\},\nonumber\\
A_i(t+\delta t,\vec{x})&=&A_i(t,\vec{x})-\delta t E_i(t+\frac{\delta t}{2},\vec{x})
,\nonumber\\
f_i^{(n)}(t+\delta t,\vec{x})&=&f_i^{(n)}(t,\vec{x})+
\delta t F_i^{(n)}(t+\frac{\delta t}{2},\vec{x})
,\nonumber\\
\theta^{(n)}(t+\delta t,\vec{x})&=&\theta^{(n)}(t,\vec{x})+
\delta t \Pi^{(n)}(t+\frac{\delta t}{2},\vec{x})
,
\labx{equ:eomLatt}
\end{eqnarray}
where we have used shorthand notations
\begin{eqnarray}
\Delta^\pm_i\phi(\vec{x})&=&
\pm\left(\phi(\vec{x}\pm\hat{i})-\phi(\vec{x})\right),\nonumber\\
A_{ij}(\vec{x})&=&
\Delta^+_iA_j(\vec{x})-\Delta^+_jA_i(\vec{x}),\nonumber\\
\tilde{\nabla}^2\phi(\vec{x})&=&
\frac{1}{a^2}\sum_i\left(\phi(t,\vec{x}+\hat{i})
-2\phi(t,\vec{x})+\phi(t,\vec{x}-\hat{i})
\right).
\end{eqnarray}

\bibliographystyle{prsty}
\bibliography{htlsimu}

\begin{figure}
\begin{center}
\input green.pstex_t
\end{center}
\caption{Analytic structure of the propagator (\ref{equ:prop}).
The filled circles are oscillatory poles, the thick line is the
branch cut, and the arrows show how imaginary poles at $\omega=\pm i\gamma_L$,
depicted by open circles,
can be found by analytically continuing the propagator through the
branch cut.}
\label{fig:prop}
\end{figure}

\begin{table}
\caption{The approximate
number of degrees of freedom needed in various approaches
to simulate a mode with momentum $k$ reliably for time $t\gg 1/k$.}
\begin{tabular}{ll}
Approach&Degrees of freedom\\
\hline
Legendre polynomials&$4kt$\\
Spherical harmonics&$(kt)^2$\\
Discrete velocities&$\frac{24}{\pi^2}(kt)^2$
\end{tabular}
\label{tab:simtimes}
\end{table}

\end{document}